# A Quantitative Physical Model of the Snow Crystal Morphology Diagram

Kenneth G. Libbrecht

Department of Physics
California Institute of Technology
Pasadena, California 91125
kgl@caltech.edu

**Abstract.** I describe a semi-empirical molecular model of the surface attachment kinetics governing ice crystal growth from water vapor as a function of temperature, supersaturation, and crystal mesostructure. An important new hypothesis in this model is surface-energy-driven molecular diffusion enabled by a leaky Ehrlich-Schwoebel barrier. The proposed surface-diffusion behavior is sensitive to facet width and surface premelting, yielding structure-dependent attachment kinetics with a complex temperature dependence. By incorporating several reasonable assumptions regarding the surface premelting behavior on basal and prism facets, this model can explain the overarching features of the snow crystal morphology diagram, which has been an enduring scientific puzzle for nearly 75 years.

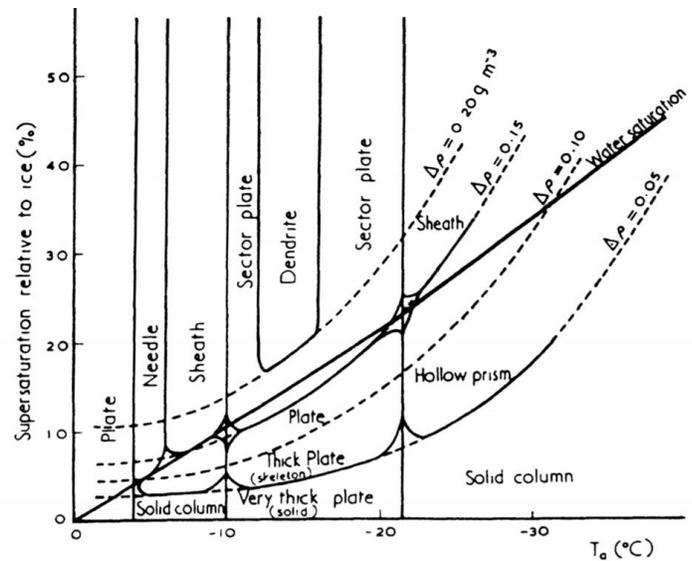

Figure 1: One version of a snow crystal morphology diagram, indicating the morphology of snow crystals growing in air as a function of temperature and water vapor supersaturation (from [1990Yok]).

## 1. The Snow Crystal Morphology Diagram

Beginning about 75 years ago, Hanajima and Nakaya presented a series of laboratory investigations describing how the morphology of snow crystals growing in air varied with temperature and supersaturation [1942Han, 1944Han, 1954Nak, 1958Nak]. This pioneering work was subsequently reproduced and extended by numerous authors [1958Hal, 1961Kob, 1990Yok, 2009Bai, 2012Bai] and Figure 1 illustrates one example of what has become known as the *Snow Crystal Morphology diagram*, or sometimes the *Nakaya diagram*. Different authors have reported some variation in the details seen in Figure 1, but several overarching trends have been consistent over time. In particular, snow crystal morphologies generally exhibit an overall plate-like behavior near -2 C, switching to slender columnar forms near -5 C, followed by especially thin plate-like structures near -15 C, and columnar crystals again around -30 C. Moreover, there is a general trend toward greater morphological complexity as the supersaturation increases; simple faceted prisms are the norm at low supersaturations, while dendritic structures appear at higher supersaturations.

The snow crystal morphology diagram has become something of a *Rosetta Stone* for snowflakes, as it provides at least a qualitative explanation for the great variety of snow crystal types that can be found in nature, as illustrated in Figure 2. For example, large stellar snow crystals appear only when the cloud temperatures are in a narrow range around -15 C, while slender needle-like forms appear only around -5 C. A capped column (for example, (f) in Figure 2) first grows as a columnar form near -5 C, and then the growth turns plate-like as the cloud rises (often in a cold front) and its temperature drops below -10 C.

Even after many decades, the snow crystal morphology diagram has remained largely qualitative in nature, based mainly on morphological observations of laboratory-grown snow crystals. This situation is slowly changing, however, as new experimental techniques are allowing precise measurements of snow crystal growth rates as well as morphologies in well-controlled environmental conditions, from well-characterized initial seed crystals, including complex morphologies [2014Lib1, 2015Lib3, 2019Lib]. Moreover, computational models of diffusion-limited solidification are beginning to reproduce realistic snow crystal structures using physically reasonable physical inputs [2014Kel, 2013Kel, 2009Gra, 2008Lib]. As rapid progress is being made on both the experimental and computational fronts, it is becoming feasible, at least in principle, to develop a comprehensive model of ice growth dynamics that can reproduce the full menagerie of observed snow crystal morphologies.

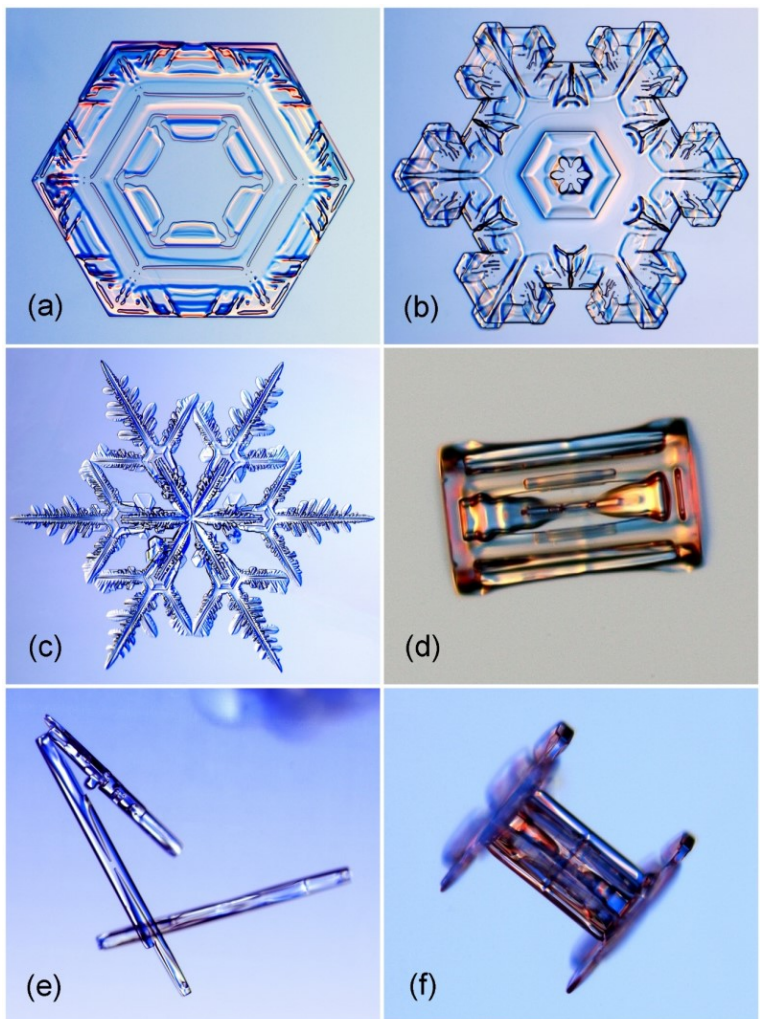

**Figure 2: Natural snow crystals can be found in a great variety of morphological types, and these examples illustrate (a) a simple plate, (b) a stellar plate, (c) a stellar dendrite, (d) a stout column, (e) several slender columns, and (f) a capped column (from [2005Lib]).**

The primary impediment limiting progress toward the long-sought goal of creating realistic computational snow crystals is a suitable understanding of the surface attachment kinetics, describing the rate at which impinging water vapor molecules attach to the ice surface under different conditions. The physics underlying the diffusion-limited transport of water vapor molecules through the air is well understood from statistical mechanics, and relatively minor physical influences like the Gibbs-Thomson (surface energy) effect and latent heating effects are also well understood and calculable. In contrast, the attachment kinetics derives from subtle and complex molecular interactions at the ice crystal surface, for which the fundamental many-body physics remains largely unknown. Even at a rudimentary, qualitative level, there is much about the attachment kinetics at the ice/vapor interface that remains mysterious. Developing even a crude quantitative model of the attachment kinetics has therefore remained a remarkably difficult problem. In the discussion below, I report what I believe is important progress toward developing a comprehensive physical model of the ice/vapor attachment kinetics, thereby explaining many previously enigmatic aspects of the snow crystal morphology diagram.

## 2. Quantitative Snow Crystal Growth data

Understanding the ice/vapor attachment kinetics from molecular first principles is far beyond the state-of-the-art in many-body molecular physics, including modern molecular-dynamics simulations, so we must rely on experimental measurements of snow-crystal growth rates to guide the discussion. Understanding the published data is somewhat challenging, however, as many reported measurements were adversely affected, sometimes strongly so, by systematic errors arising from diffusion effects, thermal effects, and substrate interactions [2004Lib, 2015Lib].

In this section, I describe what I believe are the most reliable measurements to date, and especially those measurements that most pertain to developing a model of the ice/vapor attachment kinetics. Choosing which data to include, and which to reject, is not a foolproof task, but some judgement calls are necessary when faced with the considerable inconsistencies found between different measurements reported in the scientific literature. Fortunately, the model described below suggests a large number of targeted follow-up experiments that can be applied to test its overall validity.

### Basic Theory

A quantitative discussion of ice growth rates from water vapor necessarily begins with the Hertz-Knudsen relation [1882Her, 1915Knu, 1996Sai, 1990Yok, 2005Lib, 2017Lib, 2019Lib], which we write as

$$v_n = \alpha v_{kin} \sigma_{surf} \quad (1)$$

where $v_n$ is the crystal growth velocity normal to a growing surface, $\alpha$ is a dimensionless *attachment coefficient*, $\sigma_{surf} = (c_{surf} - c_{sat})/c_{sat}$ is the water vapor supersaturation at the surface, $c_{surf}$ is the water-vapor number density just above the surface, $c_{sat} = c_{sat}(T)$ is the saturated number density above a surface in equilibrium at temperature $T$, and

$$v_{kin} = \frac{c_{sat}}{c_{ice}} \sqrt{\frac{kT}{2\pi m_{mol}}} \quad (2)$$

is the *kinetic velocity*, in which $m_{mol}$ is the mass of a water molecule, $c_{ice} = \rho_{ice}/m_{mol}$ is the number density of ice, and $\rho_{ice}$ is the mass density of ice. Values of all these quantities as a function of temperature can be found in [2019Lib].

In principle, Equation 1 can be precisely applied only for the case of a semi-infinite planar surface. In this case, the Hertz-Knudsen relation describes a simple change of variables, transforming $v_n(\sigma_{surf})$ (here at fixed temperature) into the new function $\alpha(\sigma_{surf})$. The statistical mechanics of ideal gases has been applied in this transformation so that the value of $\alpha(\sigma_{surf})$ must lie between 0 and 1, and the attachment coefficient can therefore be thought of as a sticking probability for impinging water vapor molecules. On a molecularly rough ice surface, vapor molecules typically become incorporated into the crystal lattice immediately upon impact, yielding $\alpha = 1$. The experimental evidence suggests that $\alpha \approx 1$ is a good approximation for molecularly rough ice surfaces and for liquid water surfaces, although this is not known with absolute certainty.

If a surface is not infinite in extent, it is possible for surface diffusion to produce an effective attachment coefficient that is substantially greater than unity. One famous example is the growth of mercury whiskers from vapor [1955Sea], where surface diffusion carries adatoms from the whisker sides to the whisker tip, where they readily attach at a crystal dislocation, yielding $\alpha \gg 1$ on the tip. It appears that snow crystals are typically effectively dislocation-free, so the mercury-whisker mechanism will not factor into our story here. But surface diffusion will play an important role in the discussion below, when I describe narrow facet surfaces.

On infinite faceted surfaces, impinging vapor molecules often reside on the surface temporarily (where they are called admolecules), but soon return to the vapor phase from thermal fluctuations, before they become incorporated into the ice lattice. Put another way, the admolecules on facet surfaces are distinguishable from ice lattice molecules, in contrast to the rough-surface case. For this common situation, the growth of faceted surfaces exhibits $\alpha < 1$.

On the ice basal and prism facets, growth is typically limited by the nucleation of new island terraces from surface admolecules. This 2D nucleation process has been quite well studied, and the underlying theory is described in detail in essentially all textbooks on crystal growth [e.g. 1996Sai, 1999Pim, 2002Mut]. Jumping straight to the result, theory yields an attachment coefficient that can be written [1996Sai]

$$\alpha(\sigma_{surf}) = Ae^{-\sigma_0/\sigma_{surf}} \quad (3)$$

for the growth of a faceted surface at a fixed temperature, where $A = A(\sigma_{surf})$ and $\sigma_0$ are dimensionless parameters, with

$$\sigma_0(T) = \frac{S\beta^2 a^2}{(kT)^2} \quad (4)$$

where $\beta(T)$ is the terrace step energy (the amount of energy needed to create the edge, or step, of a molecular terrace on a crystal facet), $a$ is the terrace thickness, and $S \approx 1$ is a dimensionless geometrical factor that absorbs a number of small theoretical factors [2011Lib2, 2013Lib].

The 1D step energies $\beta_{basal}$ and $\beta_{prism}$ on the primary ice facets are fundamental equilibrium quantities of these surfaces, analogous to the 2D surface energies and the 3D bulk energy (latent heat) of the ice crystal. To date, the best way to measure the step energies is by measuring $\sigma_{0,basal}$ and $\sigma_{0,prism}$, as I describe below [2013Lib].

A great deal of surface physics is wrapped up in the parameter $A(\sigma_{surf})$, but I will ignore this in the present discussion of ice growth behaviors. When the growth rate is high, island terraces nucleate quite readily, yielding so many growing terraces that the surface becomes effectively rough. At high $\sigma_{surf}$, therefore, it often is adequate to assume $A \approx 1$. At low $\sigma_{surf}$, the exponential factor so dominates the growth behavior that again it is often adequate to assume $A \approx 1$.

In [2013Lib], the authors presented substantial experimental evidence indicating $A < 1$ for prism facets when the growth temperature is above -10 C, and this adds additional complication to the theoretical discussion. For the present, however, I will put off any attempt to incorporate these data into the model, focusing instead on changes in the primary nucleation dynamics, as described in detail below. This simplification to a model with $A \approx 1$ presents an cleaner theoretical picture, and I believe it is sufficient for the phenomenon examined below. Thus I will proceed by assuming $A \approx 1$ for all cases of nucleation-limited ice/vapor growth throughout the remainder of this paper.

## ICE GROWTH IN AIR

One of the best methods for measuring snow-crystal growth rates in air is by observing small crystals that have fallen freely through an experimental chamber with a known supersaturation. Figure 3 shows one example depicting a series of measurements at different temperatures after 200 seconds of growth in air with $\sigma_\infty \approx \sigma_{water}$, which is the supersaturation of metastable liquid water relative to ice at a fixed temperature [1974Yam, 1987Kob]. Note that the variation in aspect ratio (from plate-like to columnar) is in good agreement with the snow-crystal morphology diagram. Moreover, with quantitative data like these, one can begin to extract measurements of $\alpha_{basal}$ and $\alpha_{prism}$ from the data as functions of $T$ and $\sigma_{surf}$. Additional free-fall growth measurements in air can be found in [1976Gon, 1982Gon, 2008Lib1, 2009Lib].

Unfortunately, ice growth measurements in normal air tend to be limited to providing information only about the aspect ratio $\alpha_{basal}/\alpha_{prism}$ under different conditions. They cannot easily yield accurate information about $\alpha_{basal}$ or $\alpha_{prism}$ individually because ice growth in air is strongly limited by particle diffusion, which means that $\sigma_{surf}$ is typically much smaller than $\sigma_\infty$, the latter being the supersaturation far from the growing crystal. Because it is generally impractical to measure $\sigma_{surf}$ directly, and it is quite difficult to model diffusion effects accurately as well, the resulting experimental uncertainties in $\sigma_{surf}$ are so high that $\alpha$ cannot be extracted from much of the available ice growth data with a meaningful accuracy [2005Lib, 2017Lib].

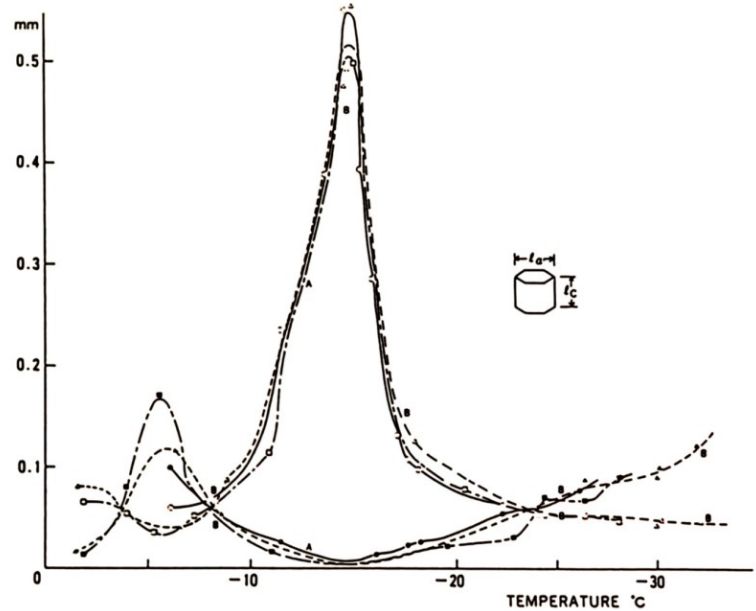

**Figure 3: Measurements of the diameters and thicknesses of snow crystals after growing in air for 200 seconds in a free-fall cloud chamber filled with air and liquid water droplets [1974Yam, 1987Kob]. Note that the variation in crystal aspect ratio as a function of temperature matches expectations from the snow crystal morphology diagram.**

Diffusion effects in air become somewhat manageable when the crystal size is exceedingly small (no more than a few microns) and/or $\alpha$ is exceedingly small, but this presents a rather small region of parameter space. There is certainly some useful information to be gleaned from growth measurements in air [1972Lam, 1976Gon, 1982Gon, 1984Cho1, 1998Nel, 2004Bai, 2008Lib1, 2009Lib], but an in-depth analysis of the results is beyond the scope of this paper. For our present discussion, the main take-away from these measurements is that the $\alpha_{basal}/\alpha_{prism}$ aspect ratios for small plate-like and columnar crystals are generally in good agreement with the morphology diagram.

Another interesting result from [2009Lib], however, is that the aspect ratios of small crystals grown at -5 C and -10 C tend to become more extreme (i.e., farther from unity) as the supersaturation is reduced, at least over a limited range in $\sigma_\infty$. This result is consistent with expectations from nucleation limited growth, as the aspect ratio would be roughly given by

$$\frac{\alpha_{basal}}{\alpha_{prism}} \approx \frac{e^{-\sigma_{0,basal}/\sigma_{surf}}}{e^{-\sigma_{0,prism}/\sigma_{surf}}}$$
$$\approx e^{-\Delta\sigma_0/\sigma_{surf}} \quad (5)$$

because $\sigma_{surf}$ is nearly the same for both facets on a small ice prism.

Regardless of the sign of $\Delta\sigma_0$, the aspect ratio becomes further from unity at lower $\sigma_{surf}$, and this trend has been observed at -5 C and -10 C [2009Lib]. At extremely low $\sigma_{surf}$, however, one expects that surface-energy effects will inhibit the growth of thin plate edges or slender needle tips, and this is likely why the aspect ratios in Figure 1 tend toward unity at the lowest supersaturations.

## Ice Growth in near Vacuum

Measuring ice growth rates in near-vacuum conditions presents a distinct advantage over observations in normal air, as diffusion effects are much reduced as the background gas pressure is reduced. Although it can be exceedingly difficult to determine $\sigma_{surf}$ on a growing ice crystal in air, one can assume $\sigma_{surf} \approx \sigma_\infty$ if the background gas pressure (not including water vapor) is sufficiently low. Moreover, $\sigma_\infty$ is typically an experimentally determined parameter, whereas measuring $\sigma_{surf}$ directly is not practical. Near-vacuum measurements are thus much better suited for investigating the attachment kinetics, as can be demonstrated using the analytical theory for the growth of a spherical crystal [2005Lib, 2017Lib, 2019Lib].

Libbrecht and Rickerby [2013Lib] presented an extensive series of measurements of $\alpha_{basal}(\sigma_{surf},T)$ and $\alpha_{prism}(\sigma_{surf},T)$ over the temperature range $-2\,C > T > -40\,C$, including a careful analysis of systematic effects arising from particle diffusion, crystal heating, and substrate interactions. Notably, their basal-growth data provided unambiguous evidence for nucleation-limited growth with $A \approx 1$ over the entire temperature range. The prism-growth data were noisier, and suffered from larger systematics, but there again it appears that the data are consistent with nucleation-limited growth with $A \approx 1$ over the full temperature range investigated.

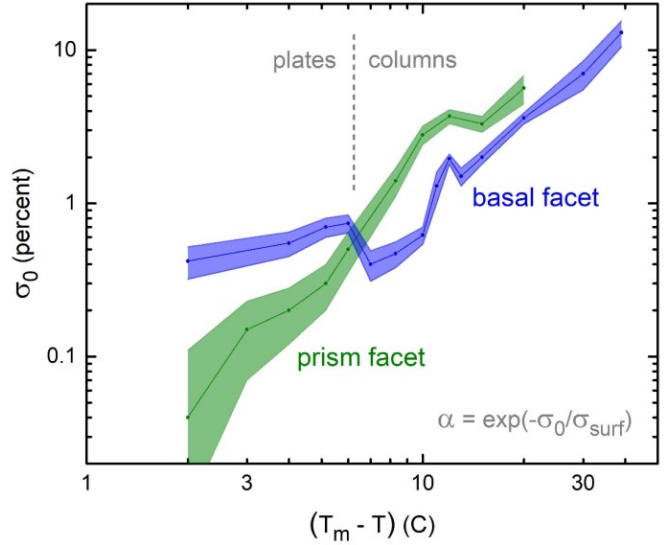

**Figure 4: Measurements of ice growth in near-vacuum are well represented by $\alpha(\sigma_{surf}) = exp(-\sigma_0/\sigma_{surf})$ with $\sigma_{0,basal}(T)$ and $\sigma_{0,prism}(T)$ shown here [2013Lib]. Taken at face value, these data suggest that plate-like crystals should grow at temperatures above $T \approx -6\,C$, while columnar crystals should grow below this temperature. Clearly this behavior is strongly inconsistent with the normal snow-crystal morphology diagram. The crystal-growth model presented in this paper provides an explanation for this striking difference.**

Figure 4 shows these measurements of $\sigma_{0,basal}$ and $\sigma_{0,prism}$, assuming that the attachment coefficient can be described by $\alpha(\sigma_{surf}) \approx \exp(-\sigma_0/\sigma_{surf})$ for both facets. Using these data, we see that the growth aspect ratio $\alpha_{basal}/\alpha_{prism}$ is less than unity for all $\sigma_{surf}$ when $T > -6\,C$, suggesting plate-like growth in this region. Similarly, the data suggest columnar growth when $T < -6\,C$, as indicated in Figure 4. This suggested behavior is much different from what is seen in the morphology diagram, thereby presenting important information for understanding the underlying attachment kinetics. The main focus of this paper is to reconcile these measurements with the morphology diagram, as I describe below.

The experimental data in Figure 4 can be converted to measurements of the terrace step energies on the basal and prism facets [2013Lib], and doing so provides some useful insights into the molecular structures of these surfaces [2017Lib, 2019Lib]. At the lowest temperatures shown, the ice surface can be thought of using the usual cartoon image of a crystal surface, where a terrace step is static and abrupt, being essentially one molecule wide. As the temperature increases, however, thermal fluctuations and surface relaxation cause the terrace edges to become more dynamic and more diffuse, so a step becomes effectively broader and gentler. In this simplistic geometrical picture, a more gradual step means a lower step energy, so the step energy, and thus $\sigma_0$, decreases with increasing temperature.

On the basal surface, the step energy decreases with increasing temperature until $\sigma_0$ apparently reaches a plateau as $T \to 0\,C$. Surface premelting becomes prominent near the melting point, yielding a thick quasi-liquid-layer (QLL) in the process. As the QLL develops, the ice/vapor step energy transforms into an ice/QLL step energy, which tends toward the ice/water step energy as $T \to 0\,C$. Thus we see a correspondence between ice growth from vapor and ice growth from water as we approach the triple point [2014Lib]. In this picture, the nonzero plateau seen in $\sigma_{0,basal}$ reflects basal faceting in ice/water solidification, while the fact that $\sigma_{0,prism} \to 0$ as $T \to 0\,C$ reflects the lack of prism faceting when liquid water freezes.

We are still far from a full physical understanding of surface premelting and ice crystal growth as a function of temperature, and it is not yet possible to calculate step energies at any temperature on any ice surface. But the overall trends in the step energy behavior, as indicated by the $\sigma_0(T)$ measurements in Figure 4, can at least be qualitatively understood from this basic physical picture.

## Previous Attempts to Explain the Morphology Diagram

Laboratory investigations spanning many decades have motivated some previous attempts to create a physical model of ice/vapor growth that can explain the overall characteristics of the morphology diagram. In an early attempt in 1958, Hallett and Mason presented a series of measurements of the diffusion length of admolecules on faceted surfaces, revealing a non-monotonic temperature behavior that could explain the temperature-dependent transitions between plate-like and columnar growth [1958Hal, 1971Mas]. This semi-empirical model has largely been abandoned, however, as it has not been supported by subsequent experimental investigations. The model did not incorporate nucleation-limited growth, with its exponential dependence on $\sigma_{surf}$, and it provides no explanation for the differences between growth in air and at low pressures. Moreover, the early measurements were strongly affected by particle diffusion effects in air that are difficult to quantify, and it appears likely that the measurements did not accurately measure surface diffusion lengths.

Kuroda and Lacmann developed a much different model in the 1980s, focusing on changes in the ice surface structure arising from the development of surface premelting [1982Kur, 1984Kur, 1984Kur1, 1987Kob]. This model was a substantial step forward in that it incorporated surface premelting as a primary driver of temperature-dependent changes in growth behavior. However, it did not include nucleation-limited growth, nor does it provide any means of explaining the profound differences between growth in air and in a near-vacuum environment. Thus, although the Kuroda-Lacmann model has interesting components, its primary features do not appear to reflect the actual physical processes involved in snow crystal growth, and the model has not been well supported by subsequent experimental investigations.

Another problematic issue with both these early models is that they were largely qualitative in nature, designed to explain a morphology diagram that was itself qualitative. As such, the models were poorly suited for making quantitative predictions regarding ice growth rates under different laboratory conditions. In contrast, the model presented below specifies the numerical values of $\alpha_{basal}(\sigma_{surf}, T)$ and $\alpha_{prism}(\sigma_{surf}, T)$ over a broad range of experimentally accessible conditions, suggesting many future experimental investigations that can examine its overall validity.

## 3. Structure Dependent Attachment Kinetics

The next step in our discussion is to consider finite faceted surfaces, meaning surfaces that are not infinite in extent. What I hope to show here is that terrace nucleation on finite faceted surfaces can differ substantial from that on infinite surfaces, and does so in the case of snow crystal growth. I introduced this concept some years ago [2003Lib1], calling it *Structure Dependent Attachment Kinetics*, or *SDAK*. This immediately led to some confusion, as people often assumed that "structure" here means the molecular structure of the crystalline surface itself. This is not true; instead I am referring to the mesoscopic structure associated with non-infinite surfaces, such as a sharp needle tip or the narrow edge of a thin plate. Since [2003Lib1], I have been arguing that some kind of SDAK effect – of previously unknown physical origin – was necessary to explain a variety of snow-crystal growth observations. In this paper, I present what I believe is a plausible physical mechanism underlying the SDAK phenomenon. Moreover, I believe that this mechanism provides a key component in a comprehensive model of the ice/vapor attachment kinetics, and thus explains the dominant temperature-dependent features observed in the snow crystal morphology diagram.

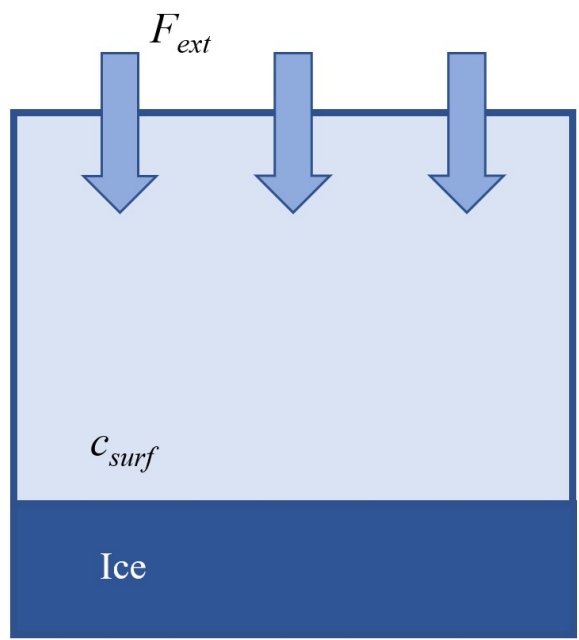

**Figure 5: A thought experiment in which a faceted ice surface is in contact with a water vapor region having a number density $c_{surf}$. In addition to the ice and the normal thermal vapor above, an external flux $F_{ext}$ of water vapor molecules is applied from above. We assume no interaction between the external flux and the vapor region, so the entire flux $F_{ext}$ strikes the ice surface, and also $c_{surf}$ is unaffected by the presence of the external flux.**

### Artificially Lowering the Nucleation Barrier

As a prelude to looking at finite surfaces, consider first the thought experiment illustrated in Figure 5. In this experiment, we start with a faceted ice surface in contact with a water-vapor reservoir with number density $c_{surf}$, and we further assume that the facet surface has an infinite horizontal extent. In addition, we impose an external flux of water vapor molecules with magnitude $F_{ext}$. In Figure 5, we depict this as a beam of water vapor molecules impinging on the surface from some external source, but its actual origin is not important.

Setting $F_{ext} = 0$ to start, the conventional flux of water molecules impinging on the ice surface can be written as $F_{in} = c_{surf}\langle v \rangle$, where $\langle v \rangle$ is an appropriate average over the thermal distribution of gas molecules. As is typical in crystal-growth theory, we assume that these impinging molecules initially become loosely attached to the faceted surface as admolecules, but thermal fluctuations cause isolated admolecules to leave the surface after residing an average time $\tau$. We assume that the facet growth is governed by 2D terrace nucleation, so only island terraces made from numerous admolecules will reside significantly longer than $\tau$. Neglecting the presence of island terraces, the flux of admolecules leaving the surface can be written $F_{out} = \rho_{surf}/\tau$, where $\rho_{surf}$ is the admolecule surface density. The balance of inward and outward fluxes, $F_{out} = F_{in}$, yields a quasi-equilibrium surface density given by $\rho_{surf} = c_{surf}\langle v \rangle \tau$. Again with $F_{ext} = 0$, we see that large island terraces will neither grow nor shrink when $\rho_{surf} = \rho_{eq} = c_{sat}\langle v \rangle \tau$, (as this implies $\sigma_{surf} = 0$), while smaller terraces will tend to shrink via the 2D Gibbs-Thomson effect. So far this discussion is typical of the standard theory of 2D nucleation-limited growth, as described in crystal-growth textbooks. Thus the attachment coefficient can be written in terms of the admolecule density as

$$\alpha = A e^{-\sigma_0/\sigma_{surf}} \quad (6)$$

where $\sigma_{surf}$ can be written as

$$\sigma_{surf} = \frac{\rho_{surf} - \rho_{eq}}{\rho_{eq}} \quad (7)$$

If we now take $F_{ext} > 0$, this increases the total incident flux to $F_{in} = c_{surf}\langle v \rangle + F_{ext}$, and this increases the surface admolecule density to

$$\rho'_{surf} = (c_{surf}\langle v \rangle + F_{ext})\tau \quad (8)$$

In turn, the increased admolecule surface density means that island terraces will grow more readily than when $F_{ext} = 0$. And, as one would expect, this also means that the size of a critical nucleus (in 2D nucleation theory) will become smaller as $F_{ext}$ increases. Put another way, the nucleation rate will increase when $F_{ext}$ increases, and thus so will $\alpha$. Rewriting Equation 6 gives the attachment coefficient

$$\alpha = A\exp\left(\frac{\sigma_0 \rho_{eq}}{\rho'_{surf} - \rho_{eq}}\right) \quad (9)$$

$$= A\exp\left(\frac{\sigma_0}{\sigma_{surf}(1 + F_{ext}/\sigma_{surf}c_{sat}\langle v \rangle)}\right)$$

which reduces to Equation 6 when $F_{ext} = 0$, as it must.

Next, for reasons that will become clear in the next section, we will assume that $F_{ext} = 0$ in equilibrium (when $\sigma_{surf} = 0$), and we therefore write $F_{ext} = G c_{sat}\langle v \rangle \sigma_{surf}$ to satisfy this equilibrium condition automatically. The exact functional form of the dimensionless function $G(\sigma_{surf})$ is unknown, but a lowest-order Taylor expansion suggests that $G$ will likely be roughly constant in the limit of small $\sigma_{surf}$. Plugging this in gives

$$\alpha = A\exp\left(\frac{\sigma'_0}{\sigma_{surf}}\right) \quad (10)$$

where

$$\sigma'_0 = \frac{\sigma_0}{1 + G} \quad (11)$$

Note that the physics described by this discussion is quite straightforward; imposing an external flux of water molecules increases the admolecule surface density, and this brings about an increase in the terrace nucleation rate, as one would expect. For the special case where the external flux is proportional to $\sigma_{surf}$ (which we write as $F_{ext} = G c_{sat}\langle v \rangle \sigma_{surf}$), the nucleation rate goes up as if $\sigma_0$ were replaced by the smaller value $\sigma'_0$ in Equation 11. Thus

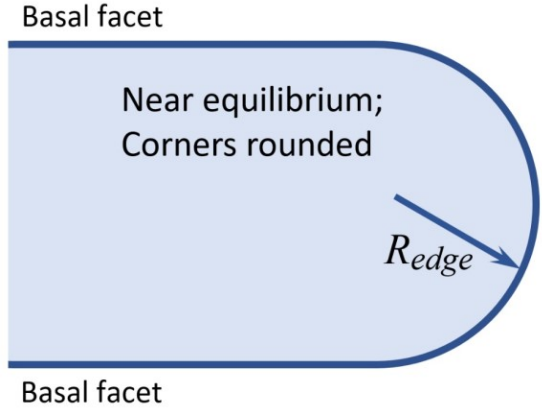

**Figure 6:** A schematic diagram showing a side view of the edge of a thin plate-like snow crystal. The rounded edge has a radius of curvature equal to $R_{edge}$, and the plate is otherwise defined by two basal facets.

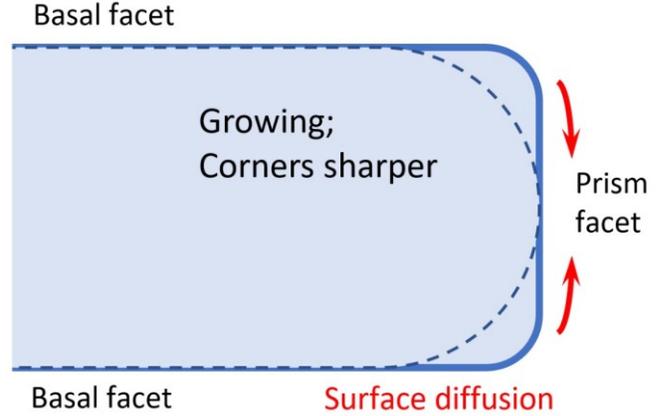

**Figure 7:** In this sketch, the plate edge in Figure 6 is subjected to a significant supersaturation, which causes the corners to sharpen. The non-equilibrium shape drives surface diffusion (red arrows) that bring a flux of admolecules to the narrow prism facet.

the external flux serves to reduce the nucleation barrier.

Essentially all aspects of the preceding discussion derive from the classical theory of 2D nucleation on faceted surfaces. Normally this theory is presented in one of two extremes: 1) where the material vapor pressure is high (like ice) and $\sigma_{surf}$ is generally small with $F_{ext} = 0$, and 2) where the material vapor pressure is essentially zero (like silicon) and an external flux $F_{ext} > 0$ is provided by chemical-vapor deposition (CVD). The discussion here simply treats the intermediate case, between these two extremes, where both the natural vapor pressure and the external flux are important for determining the 2D nucleation rate.

## Surface Diffusion on Finite Faceted Surfaces

Next consider the situation of the thin-plate snow crystal illustrated in Figure 6. To counter the Gibbs-Thomson effect at the plate edge, we immerse the crystal in a supersaturated environment with $\sigma_{surf} = d_{sv}\kappa$, where $d_{sv} = \gamma_{sv}/c_{ice}kT \approx 1\, nm$, $\gamma_{sv}$ is the solid/vapor surface energy, $\kappa \approx 1/R_{edge}$ is the surface curvature and $R_{edge}$ is the radius of curvature of the plate edge as shown in Figure 6. We assume that the radius of the disk is much greater than $R_{edge}$.

If we assume a large nucleation barrier on the basal surfaces, then the crystal shown in Figure 6 is essentially stable; it neither grows nor shrinks appreciably over short times. The small supersaturation counters the Gibbs-Thomson effect that would otherwise cause the plate edge to sublimate if $\sigma_{surf} = 0$. Implicit in this discussion is the assumption that the equilibrium shape of an ice crystal is essentially spherical, a statement that is not known with certainty, but is supported by the preponderance of the experimental data [2012Lib2].

Beginning with the stable state shown in Figure 6, we now increase the applied supersaturation, causing the crystal to rapidly transform into the slightly different shape illustrated in Figure 7. The rounded corners have now grown out until a new quasi-equilibrium state is reached, where the corners of the crystal have a reduced radius of curvature, once again balanced by the Gibbs-Thomson effect. In this situation, we have assumed large nucleation barriers on both the basal and prism surfaces, so the growth of both these surfaces is negligibly slow.

To see how this relates to real snow crystals, taking $R_{edge} \approx 1 \, \mu m$ gives $\sigma_{surf} = d_{sv}\kappa \approx 0.1$ percent in Figure 6, and this geometry could represent a thin plate at -15 C. Increasing $\sigma_{surf}$ to 1 percent, the corners sharpen until they have radii of curvature of about 0.1 $\mu m$, giving a shape much like that shown in Figure 7. Moreover, a plate edge growing at velocity $v_n$ must be experiencing $\sigma_{surf} > v_n/v_{kin}$, so 1 percent would not be atypical for thin plates growing near -15 C.

Next let us allow some surface diffusion to take place, by which I mean the transport of water molecules along the surface from one region to another. In this case, both crystals in Figures 6 and 7 would slowly transform to reduce the overall surface energy, meaning they would both slowly evolve toward spherical shapes. To simplify the discussion further, let us further assume that surface diffusion onto the basal surfaces is negligible, while diffusion onto the prism surface is not.

In Figure 6, surface diffusion restricted to the edge region would do nothing, as the edge already has a rounded, near-equilibrium shape. Thus, additional material transport would not reduce the overall surface energy (again, not allowing surface diffusion onto the basal plane). In this case, surface diffusion has essentially no effect.

In Figure 7, however, surface diffusion of molecules from the sharp corners to the flat prism plane would tend to round the edge and thereby lower the overall surface energy. Thus the nonequilibrium shape of the edge would drive surface diffusion, as indicated by the red arrows in Figure 7. This situation sets up an unusual chain of processes – water vapor first deposits onto the corners (where there is no nucleation barrier), and some of these molecules then transport onto the prism facet (where we are assuming a high nucleation barrier), and there they end up going back into the vapor phase in a time $\tau$. The high applied supersaturation drives a nonequilibrium material flow that continues as long as the externally applied supersaturation is maintained. If the prism facet has a sufficiently high nucleation barrier, then this surface-energy-driven flow occurs in the absence of any appreciable growth of the prism facet.

Now consider what happens to the admolecule density on top of the narrow prism facet shown in Figure 7. In the absence of surface diffusion, the admolecule density is determined solely by exchange with the vapor phase, as described in the previous section. But surface diffusion provides a new flux of admolecules onto the prism facet, which we can equate with the external flux $F_{ext}$ described above. We see that $F_{ext} = 0$ when $\sigma_{surf} \approx 0$, as there can be no surface-energy-driven surface diffusion in equilibrium. Thus, as described in Equation 11, the addition of surface diffusion onto the prism facet decreases the nucleation barrier by a (difficult to determine) factor of $1 + G$.

## Surface Diffusion and Surface Premelting

Most of the above discussion is based on well-understood crystal-growth physics, such as the Gibbs-Thomson effect and 2D nucleation-limited growth on faceted surfaces. The biggest unknown is the degree of surface diffusion around corners, as depicted in Figure 7. If this process is of negligible importance, then the discussion becomes irrelevant in regard to actual snow crystal growth. Our next task, therefore, is to examine whether this particular type of around-corner surface diffusion occurs, and, if so, under what circumstances.

For the case of a low-vapor-pressure crystal (e.g. silicon), a substantial Ehrlich-Schwoebel (E-S) barrier would likely strongly inhibit surface diffusion of this nature. Here again, the E-S barrier is well described in crystal-growth textbooks [e.g. 1996Sai, 1999Pim, 2002Mut], so I will not describe it here. Suffice it to say that the process depicted in Figure 7 would involve surface transport over hundreds or thousands of individual prism-facet terrace steps, and this transport would be substantially impeded if there were a high E-S barrier.

At low temperatures, say below -30 C, the ice vapor pressure is quite small, so I propose that the prism-facet E-S barrier does strongly inhibit around-corner surface diffusion in this regime. The process is impossible to calculate in any meaningful way, given our rather poor understanding of the molecular dynamics of ice surfaces, but that may change with future theoretical developments. For now, assuming a high E-S barrier at low temperatures seems like a reasonable hypothesis. In this regime, therefore, there would be no reduction in $\sigma_0$ from the SDAK mechanism described above.

Looking at the opposite extreme, the picture changes quite dramatically when surface premelting becomes an important factor in the ice surface structure. At high temperatures, when the QLL becomes especially thick, the underlying concept of the E-S barrier is itself questionable, as the ice/QLL interface begins to resemble an ice/water interface, while the E-S barrier is more of a solid/vapor phenomenon.

Beyond the E-S barrier, even the concept of an admolecule surface density begins to lose its meaning in the presence of surface premelting. The QLL itself is essentially a thick layer of disordered admolecules covering a rigid crystalline surface. The fundamental molecular processes underlying 2D nucleation theory no longer apply in this situation, and the crystal growth physics transforms into that present at an ice/water interface.

Given that the crystal growth physics at an ice/water interface is so different from that at an ice/vapor interface, I hypothesis that again the around-corner surface diffusion ideas presented above are not relevant. At especially high temperatures, when the QLL is quite thick, I propose that the nucleation barrier on the prism facet would be defined solely by the ice/QLL surface structure. In this regime, therefore, $\sigma_0$ would again be unchanged by the above SDAK mechanism.

Where it all becomes interesting is at intermediate temperatures, when surface premelting provides enough disorder to significantly reduce the E-S barrier, but the QLL thickness is still very small, perhaps of order a monolayer or less. Unfortunately, our understanding of surface premelting is especially rudimentary in this "onset" regime, so there is no way to calculate what is going on with any meaningful accuracy. Molecular dynamics simulations could be used to investigate the E-S barrier at the onset of surface premelting, but obtaining reliable results from such simulations seems unlikely with the present state-of-the-art.

Nevertheless, is seems plausible, at least to me, that some level of around-corner surface diffusion might be present in this onset regime, at temperatures just below the initial appearance of surface premelting. Moreover, it also seems plausible that the SDAK mechanism might become significant within this temperature range. If this is correct, then the change in $\sigma_0$ would look something like what is sketched in Figure 8.

As just discussed, I hypothesis that over-corner surface diffusion, and therefore the proposed SDAK mechanism, is unimportant at both the high-temperature and low-temperature ends of the plot in Figure 8. In these regions, $\sigma_0$ on a narrow facet surface is unchanged from that on a broad facet surface. The overall temperature trend is then determined by the behavior of the step energy as a function of temperature, as described previously.

The proposed "SDAK dip" in Figure 8 applies only to narrow facet surfaces, and over a fairly narrow range of temperatures just below the onset of surface premelting. On large faceted surfaces, the depth of the dip would be zero, as the SDAK mechanism applies only to surfaces near facet corners. The dip would be deepest on narrow facets closely flanked by two corners, as shown in Figures 6 and 7. The position of the dip is determined by the onset of surface premelting on the relevant facet surface (the prism facet in Figure 7), which is somewhat constrained by other measurements, but not tightly so. Neither the width nor depth of the dip structure in Figure 8 is well constrained by measurements.

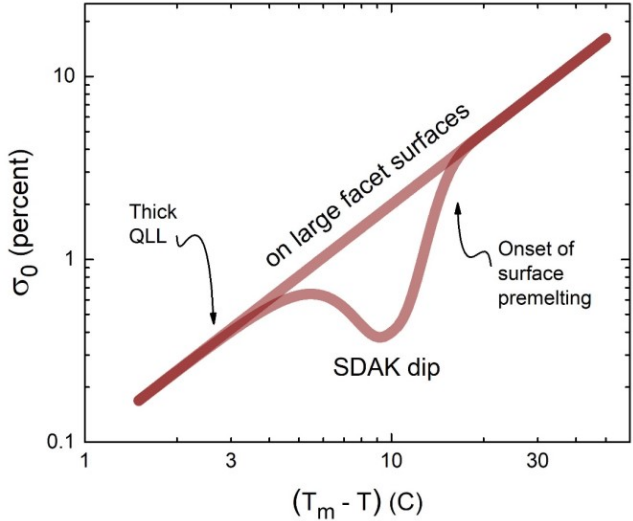

**Figure 8:** The straight line above indicates $\sigma_0(T)$ on a large faceted surface (here a generic, fictitious facet surface), roughly resembling the prism facet data shown in Figure 4. The lower curve represents the proposed SDAK mechanism. Over a limited temperature range, over-corner surface diffusion lowers $\sigma_0(T)$ compared to that on broad facets. Note that this "SDAK dip" applies only to narrow facet surfaces.

At this point it is probably useful to accept that the SDAK model proposed here is quite speculative and clearly not well constrained by solid theoretical considerations or experimental data. This, unfortunately, is the nature of the snow-crystal-growth problem. Our overarching goal here is not to present the final word on this subject, but to present a plausible physical model that could explain the snow crystal morphology diagram along with other existing experimental ice-growth measurements. Speculation of this kind can be quite useful in that it motivates further theoretical thinking and targeted experimental investigations.

Pushing forward, and recognizing that creating a model of a complex phenomenon like snow crystal growth requires some creative license, I have used existing data from a variety of sources, plus some intuition arising from years of experience with this subject, to hypothesize the SDAK dips illustrated in Figure 9. These curves have been replotted in Figure 10, this time normalizing by $\sigma_{water}$, as this better displays and contrasts the temperature-dependent growth behaviors for both facets.

A key take-away from this model is the lower panel in Figure 10, which provides a good match with the morphology diagram. We see thin plates near -2 C, columns near -5 C (with narrow basal facets, which is the case for observations in air), thin plates near -15 C (with narrow prism facets, which is the case in air), and columns again below -30 C. Of course, the placement of the SDAK dips was chosen to achieve this match with the morphology diagram, but the model makes many additional predictions that can be tested in future experimental investigations.

The SDAK idea that the attachment kinetics may depend on facet width brings a new level of richness to the snow-crystal growth problem, one that I believe has been largely overlooked to date. We have reported evidence, for example, that the high value of $\alpha_{prism}$ on thin plates near -15 C seems to be present only on narrow prism edges; broader prism facets seem to exhibit a substantially lower $\alpha_{prism}$, even under otherwise identical conditions [2015Lib2]. Similarly, columnar forms near -5 C exhibit a high $\alpha_{basal}$ only on narrow basal facets [2013Lib2]. More measurements are needed to confirm these results, but the model proposed here naturally explains a width-dependent attachment kinetics.

One especially important feature that sets the model in Figures 9 and 10 apart from all previous efforts is its *quantitative* nature. By specifying $\sigma_{0,basal}$ and $\sigma_{0,prism}$ as a function of temperature, the model specifies the numerical values of $\alpha_{basal}(\sigma_{surf}, T)$ and $\alpha_{prism}(\sigma_{surf}, T)$ over a broad range of experimentally accessible conditions. Thus the model allows for direct comparison with growth-rate measurements. Moreover, these parameters can serve as physical inputs into computational models of ice solidification, allowing investigations of complex structure formation that are again can be compared directly with laboratory experiments.

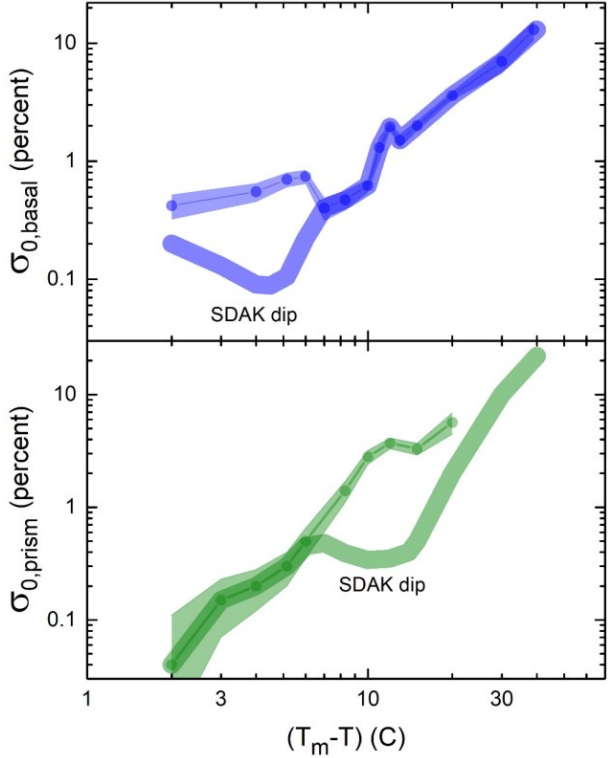
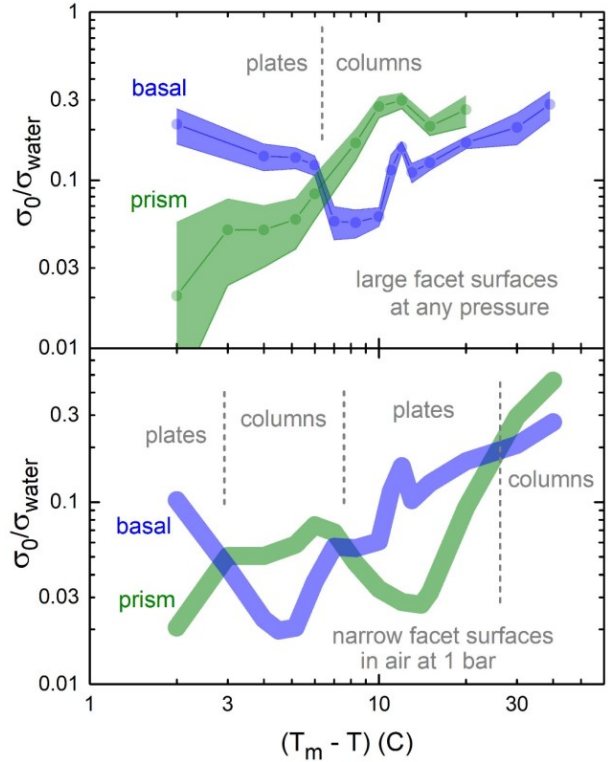

**Figure 9: Proposed values of $\sigma_{0,basal}(T)$ (top panel) and $\sigma_{0,prism}(T)$ (bottom panel) that constitute a fairly comprehensive, quantitative model of the attachment kinetics on the basal and prism facets. In each panel, the upper curves reproduce the measurements shown in Figure 4. The lower curves include SDAK dips (see Figure 8) that apply to narrow faceted surfaces.**

**Figure 10: Proposed values of $\sigma_{0,basal}(T)$ and $\sigma_{0,prism}(T)$, here normalized by $\sigma_{water}(T)$. The top panel shows values for large facet surfaces, reproducing the experimental data in Figure 4. The lower panel includes the SDAK dips and thus describes the growth of narrow facet surfaces. Because the SDAK dips apply only on the thin edges of plates near -15 C and on the thin edges of slender (usually hollow) columns near -5 C, the temperature-dependent morphologies in the lower panel bear a good resemblance to intermediate-$\sigma_\infty$ growth in the snow crystal morphology diagram.**

## Edge Sharpening Instability

Another important feature of the present SDAK model is how it can interact with diffusion-limited growth to facilitate the growth of narrow faceted structures, via a mechanism I have called the *Edge-Sharpening Instability* (ESI) [2012Lib3]. The essential idea is illustrated in Figure 11, which depicts the growth of the faceted corner of an ice crystal resting on a substrate in supersaturated air.

When the crystal grows slowly (middle sketch in Figure 11), terraces nucleate near the exposed corner of the crystal, and these terraces grow toward the facet centers to maintain the overall faceted structure of the crystal. Thus the facets have a slightly concave structure, which is the normal picture for the diffusion-limited growth of faceted crystals.

Increasing the applied supersaturation, however, brings about a break-away growth of a thin plate from the block-like pedestal, as illustrated in the right sketch in Figure 11. As the supersaturation is increased, the rate of nucleation of new prism terraces goes up, and this decreases the width of the top prism terrace. From the SDAK effect, this narrowing of the prism width decreases $\sigma_{0,prism}$ as shown in Figure 9. With a lower nucleation barrier, the terrace nucleation rate increases further, thus narrowing the width of the top terrace even more. The resulting positive feedback leads to runaway plate-like growth, which is the ESI.

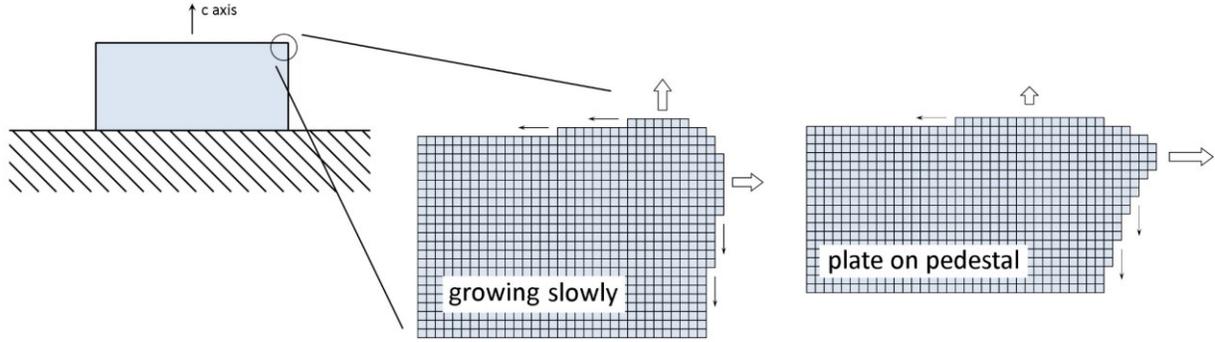

**Figure 11:** The Edge-Sharpening Instability (ESI) will promote the formation of narrow faceted features, here showing the formation of a "plate-on-pedestal" structure. When the growth rate is low (middle sketch), a normal faceted crystal develops. Above some threshold (right sketch), positive feedback leads to thin-plate formation, as $\alpha_{prism}$ reduces as the narrow prism edge develops.

## 4. Discussion

While the model proposed here is somewhat complicated and speculative, I believe it presents several new ideas that are relevant to snow-crystal growth dynamics. For many years, I have been finding experimental evidence that seems to indicate that the attachment kinetics on narrow faceted surfaces differs from the attachment kinetics on broad facets. To explain these experiments, I have suggested that some kind of SDAK mechanism is required, even though there was previously no known theoretical model that could explain this behavior. The model proposed here solves this problem, as it provides a straightforward physical mechanism that alters nucleation-limited growth on narrow facet structures, as the experiments suggest [2015Lib2, 2013Lib2]. Although there is much uncertainty in the model specifics, the overall physical picture seems quite plausible.

Importantly, this model is highly quantitative, providing clearly predicted values for $\alpha_{basal}(\sigma_{surf}, T)$ and $\alpha_{prism}(\sigma_{surf}, T)$ over a broad range of experimentally accessible conditions. The model thus suggests a host of targeted ice-growth experiments than can be used to confirm (or reject) the model's many detailed predictions, and to further refine the shapes of the SDAK dips and other model features. This model also suggests that initial conditions matter in ice-growth experiments, as does a crystal's entire growth history. The morphology diagram is only a qualitative beginning to a full understanding of ice-growth dynamics as a function of temperature, supersaturation, background-gas pressure, chemical effects, and other parameters.

Finally, the proposed model can immediately be incorporated into computational models of diffusion-limited snow crystal growth, which have been rapidly improving in recent years [2014Kel, 2013Kel, 2009Gra, 2008Lib]. An accurate, easy-to-use parameterization of the attachment kinetics has not yet been available for this purpose, so the current model suggests additional lines of investigation to better understand the formation complex snow-crystal structures that have remained unexplained for far too long.

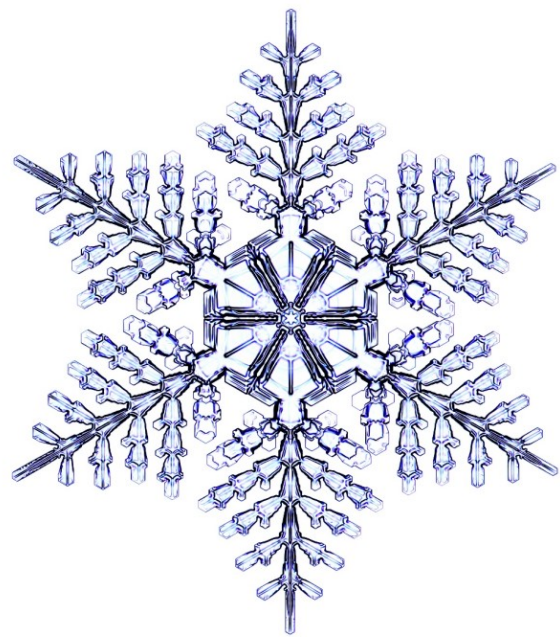